\documentclass[graybox]{svmult}

\usepackage{type1cm}        
\usepackage{makeidx}         
\usepackage{graphicx}        
\usepackage{multicol}        
\usepackage[bottom]{footmisc}

\usepackage{newtxtext}       %
\usepackage[varvw]{newtxmath}       

\makeindex             

\usepackage[T1]{fontenc}%
\usepackage[hmargin=1.5cm]{geometry}
\usepackage{graphicx}
\usepackage{dcolumn}
\usepackage{amsmath}
\usepackage{hyperref}
\usepackage {xcolor}

\begin{document} 

\title*{Monocentric or polycentric city? An empirical perspective} 

\author
{R\'emi Lemoy\orcidID{0000-0002-0396-9488}}

\institute{Rémi Lemoy \at University of Rouen, IDEES Laboratory UMR 6266 CNRS, Mont-Saint-Aignan, France, \email{remi.lemoy@univ-rouen.fr}}

\maketitle

\abstract{Do cities have just one or several centers? Studies performing radial or monocentric analyses of cities are usually criticised by researchers stating that cities are actually polycentric, and this has been well known for a long time. Reversely, when cities are studied independently of any center, other researchers will wonder how the variables of interest evolve with the distance to the center, because this distance is known to be a major determinant at the intra-urban scale.
Both monocentric and polycentric formalisms have been introduced centuries (respectively, decades) ago for the study of urban areas, and used both on the empirical and the theoretical side in different disciplines (economics, geography, complex systems, physics...). The present work performs a synthesis of both viewpoints on cities, regarding their use in the literature, and explores with data on European urban areas how some cities considered to be the most polycentric in Europe compare to more standard cities when studied through a combination of radial analysis and scaling laws.}

\section{Introduction}

Monocentric and polycentric perspectives on cities are frequently opposed in the scientific literature. A radial analysis of cities (monocentric perspective) is quite often automatically criticised by researchers stating that it has been well known for a long time that cities are polycentric -- which is certainly true. However, when researchers study cities independently of any center (polycentric perspective), some colleagues will wonder how variables evolve with the distance to the center, because this distance is known to be a major determinant at the intra-urban scale.
Both formalisms, polycentric on the one hand, radial or monocentric on the other, have been used for a long time, on the empirical as well as the theoretical side. The aim of this chapter is not to determine which point of view is right (which would be a hopeless endeavour), but rather to explore some empirical facts and understand the relations between these two formalisms and their relation to the empirical reality of cities.

Polycentrism is closely related to urban sprawl, one of the most striking phenomena appearing in the evolution of cities, since the 19th century at least. 
Many works, for instance \cite{anas98}, have described how households have been able, over time, to live further away from their jobs thanks to the decrease in transport (and commuting) prices through the availability of different means of transport:  horse, streetcar, natural combustion engine.
As cities sprawl, the distances between nearby cities decrease. Simultaneously, substructures appear, which can be represented with a gravitational mechanism \cite{li2021}. Cities grow and sprawl through the formation and emergence of secondary centers. New kinds of centrality appear besides main city centers, which can also be linked to fractal approaches (\textcolor{black}{see Chapter 8} or \cite{caruso2011}).

We focus in this chapter our empirical analysis on European cities, which we studied in previous work \cite{lemoy2020, lemoy2021} mainly through their radial (center-periphery) profiles of artificial land use -- even though we note that most of the literature studies other kinds of data. 
We study examples of polycentric urban areas, and illustrate this phenomenon using old maps compared to contemporary extensions of cities. French cities are primarily studied since historical data is readily available. Indeed, we use mainly the Cassini map, the first map of the whole kingdom of France, created by several generations of the Cassini family in the 18th century and presented in several figures here.

In the following, we start with a historical approach and a first qualitative observation of some major polycentric areas in Europe (sections \ref{history} and \ref{empirics}). Then we present a quantitative approach using radial analysis and scaling laws and a review of polycentric approaches (section \ref{models}), before a critical section, open issues and a summary.

\section{History} \label{history}

On the theoretical side, the oldest quantitative models of cities from urban economics \cite{Fujita89} and even the precursor von Thunen model \cite{vonthunen} before that have a very monocentric approach to cities, even though this body of literature has shifted to a mainly polycentric view decades ago \cite{hartwick1974, white1976, anas98}. However, it is in this disciplinary context that polycentricity was first introduced and studied, both on the empirical and theoretical side. Hence, the empirical side of the economic literature on polycentricity studies rather the location of jobs, between the Central Business District (CBD), employment subcenters  and more dispersed locations \cite{giuliano1991, mcmillen2001, arribas2014}. Indeed, in the Alonso-Muth-Mills (AMM) monocentric model, which is the standard urban economic model \cite{Fujita89}, the location of the city center is identified with the location of the (main) employment center (CBD). In the subsequent economic literature studying polycentric versions of this AMM model, several employment centers or subcenters are usually considered.

Mentioning pioneers in urban science and planning, we note that in the beginning of the 20th century, Patrick Geddes used the term ``conurbation'' to describe the process of neighbouring cities tending to mix together as they expand and sprawl \cite{geddes1915}.

\begin{figure}[!h]
\begin{center}
\includegraphics[width=\textwidth]{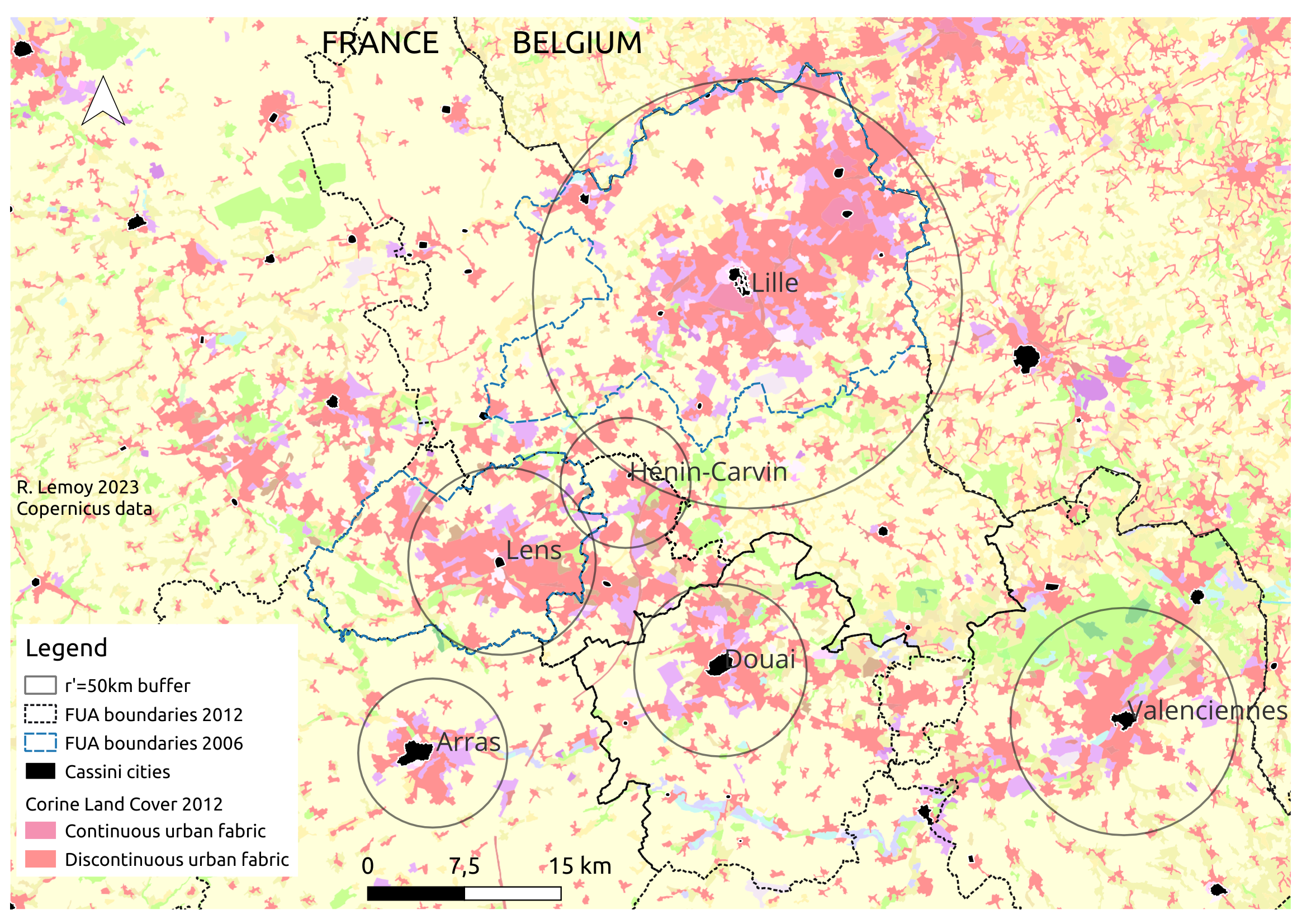}
\caption{Land use map of the Lille region FUAs (France/Belgium). The standard Corine Land Cover color scheme is used. Buffers represent a rescaled radius $r'=50$km around the center. Data from the Cassini map is overlaid, showing cities' extent in the 18th century.}
\label{lille}
\end{center}
\end{figure}

\begin{figure}[!h]
\begin{center}
\includegraphics[width=\textwidth]{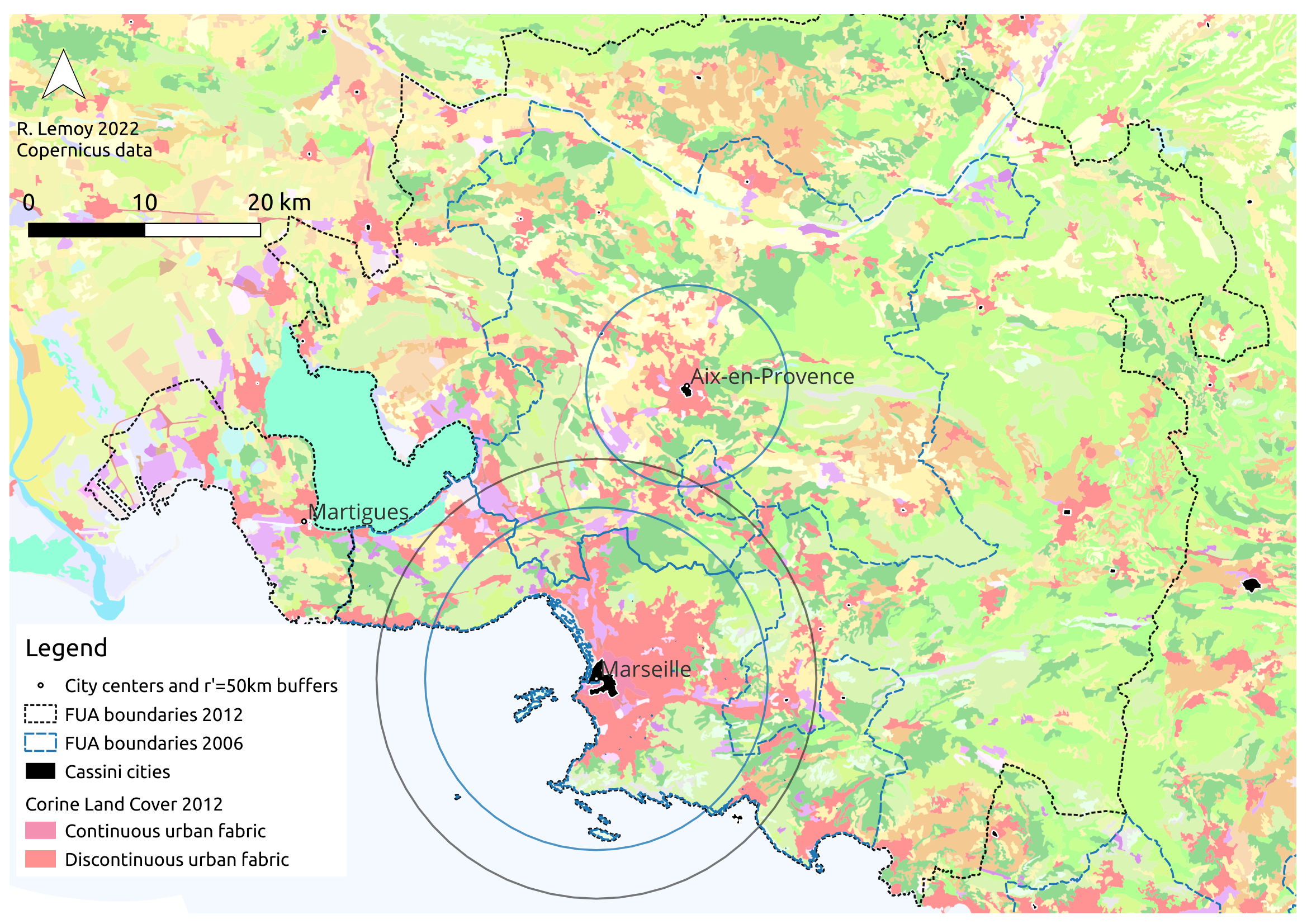}
\caption{Land use map of Marseille and Aix-en-Provence (France). The standard Corine Land Cover color scheme is used. Buffers represent a rescaled radius $r'=50$km around the center. Blue buffers correspond to 2006 data and the black one to 2012. Data from the Cassini map is overlaid, showing cities' extent in the 18th century.}
\label{marseille}
\end{center}
\end{figure}

The concept of polycentric city or polycentric urban area really emerges in the scientific literature in the 1970's - 80's, when researchers observing suburbanization (urban sprawl) and its link with transportation argue that for an efficient public transport (transit) system, it is better to have secondary centers \cite{schneider1981,gordon1985}. A positive view of polycentrism develops at this time. In terms of urban planning, the polycentricity agenda is opposed to urban sprawl. Having several dense centers yields a preservation of open space, as opposed to a continuum of low density urban space. However, it remains clear that the origin of polycentricity lies in urban expansion (and sprawl). 

More recently, the complex systems literature close to statistical physics works on the subject of polycentric cities often using rather new data generated by digital footprint \cite{roth2011, louail2014, lenormand2015}. City centers and sub-centers are linked with different kinds of activities, not just employment. Explanatory models are also proposed \cite{louf2013}.

We note also that different terms have been used in the literature to describe this phenomenon of polycentric urban areas, with different goals and varying success, among which supercity, megaregion, metropolitan region or area, urban region. We rather stick to the term ``polycentric'' here. We can also note that ``polycentrism'' and ``polycentricity'' seem to be equally used with quite similar meanings.

\section{Empirics}
\label{empirics}

\begin{figure}[!h]
\begin{center}
\includegraphics[width=\textwidth]{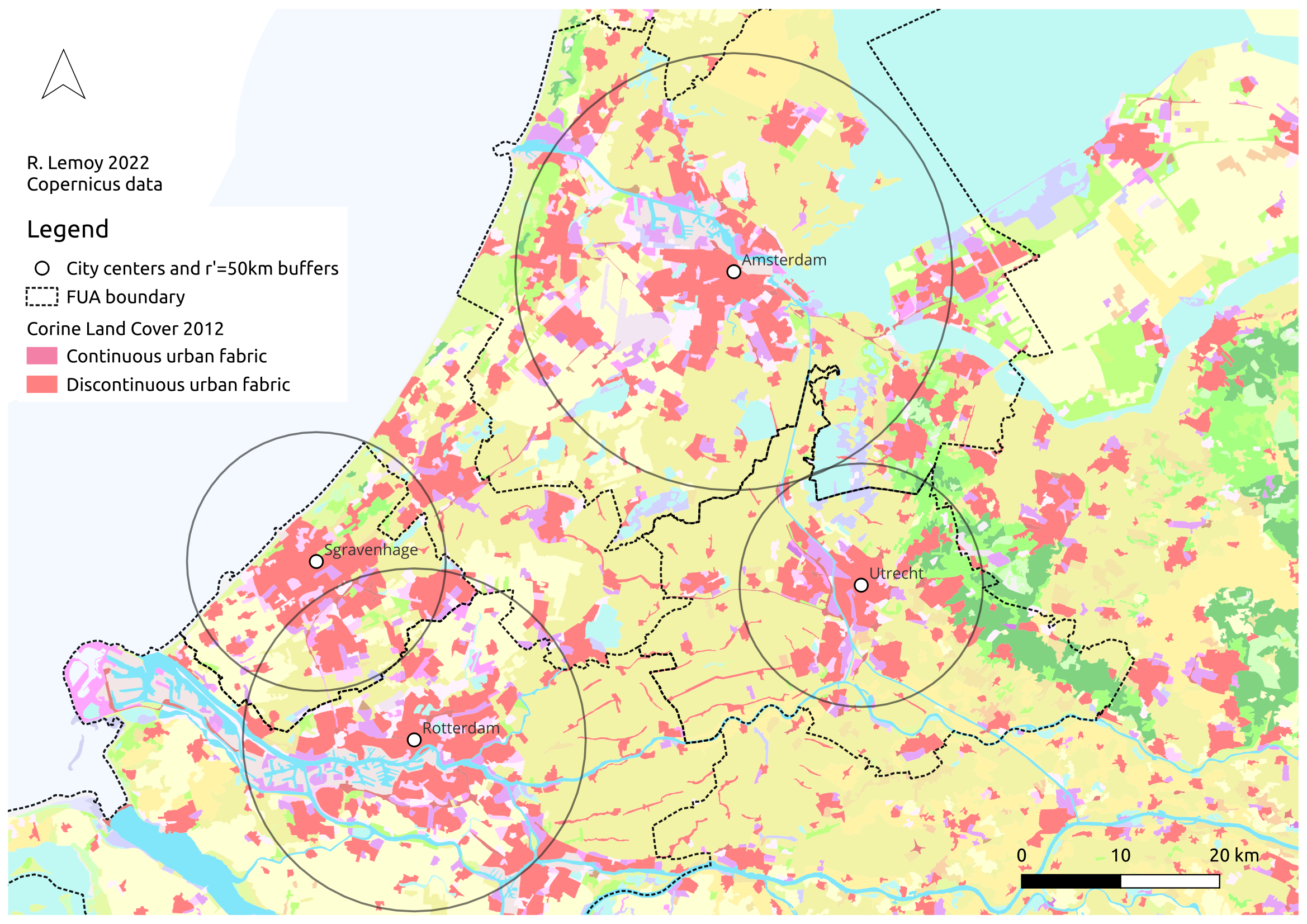}
\caption{Land use map of the Ranstad (the Netherlands). The standard Corine Land Cover color scheme is used. Buffers represent a rescaled radius $r'=50$km around the center.}
\label{ranstad}
\end{center}
\end{figure}

\begin{figure}[!h]
\begin{center}
\includegraphics[width=0.5\textwidth]{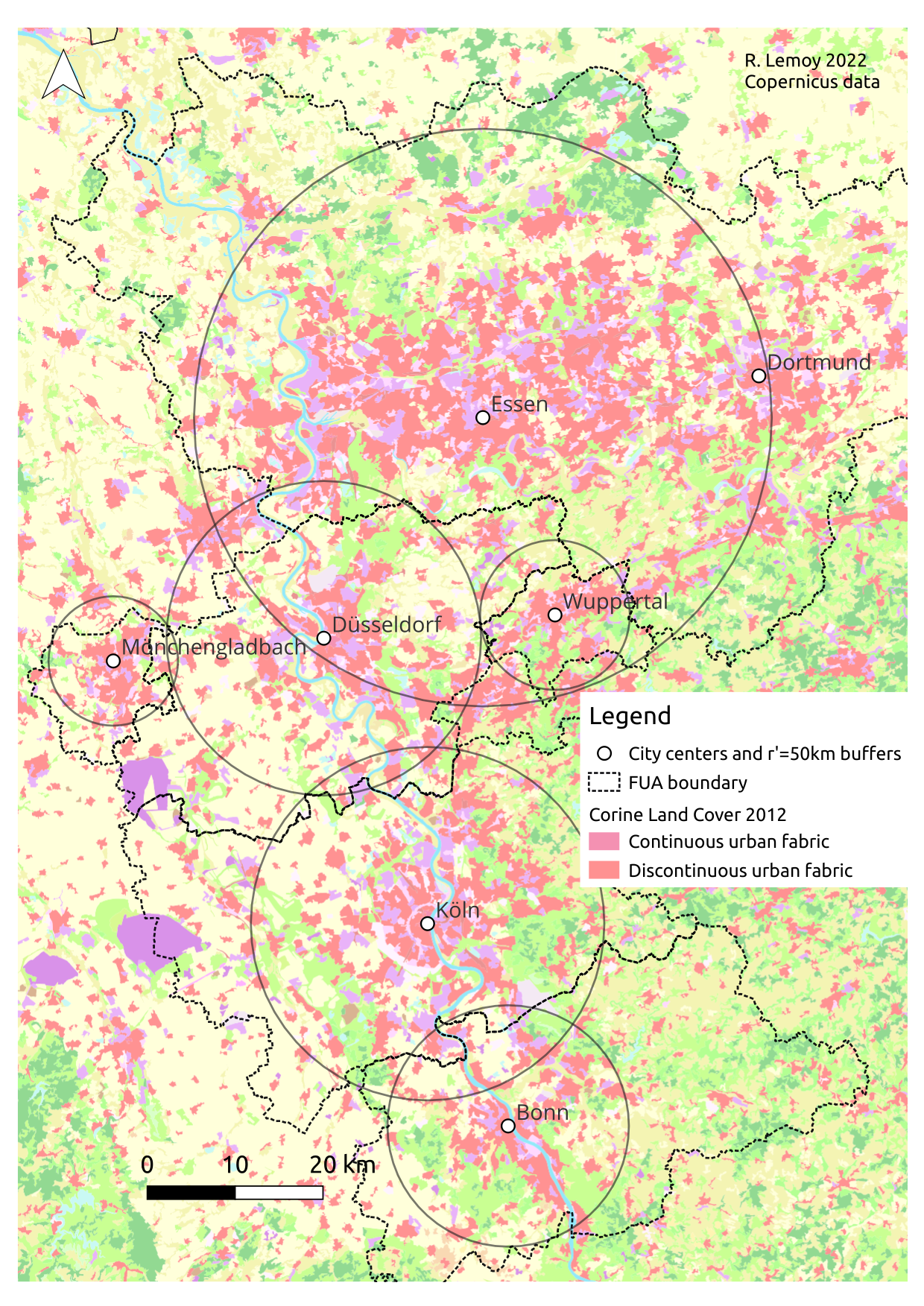}
\caption{Land use map of the Rhine-Ruhr region (Germany). The standard Corine Land Cover color scheme is used. Buffers represent a rescaled radius $r'=50$km around the center.}
\label{rhine-ruhr}
\end{center}
\end{figure}

What is a polycentric city? As the name implies, it represents a city with several centers, usually a hierarchy of centers, for instance one main center and several subcenters.
However, all cities have subcenters, since economic activities (and human activities in general) are usually clustered in certain areas of cities. Are all cities polycentric, then? To some degree, we can answer in a positive way. This is a legitimate point of view, as it is legitimate to consider that urban areas are organised, in general, around one main central point (monocentric point of view). Any city can be considered as polycentric or monocentric, depending on the way it is studied. A rather close look at local density variations yields a polycentric view, and a rather wide look at the same variables suggests a monocentric view.

Here we wish to go further, and first aim to find good examples of polycentric cities. For comparison, it is probably easier to find good examples of monocentric cities. Cities which are not too close to their neighbours and do not have special physical geography characteristics (coastline, mountainous area, ...) follow quite closely the general rule provided by radial scaling analysis \cite{lemoy2020,lemoy2021}. What would be then a good example of a polycentric city? In the view of researchers and policymakers, there seem indeed to be some cities which are specifically polycentric.

Actually, the examples of polycentric cities mentioned in the literature are usually located in very urbanised regions, which have naturally been expanding or sprawling in the last decades or centuries. 
Figures \ref{lille} and \ref{marseille} present the state of urbanization in the 18th century (using data from the Cassini map) for the Lille and Marseille areas respectively, compared with Corine Land Cover 2012 land use data.
We see on these maps that cities which were clearly distinct three centuries ago have now spread so much that the distinctions between their spatial extents are not clear anymore, forming a polycentric urban area -- which is also clearly crossing a border in the case of the Lille area. We can also see that contemporary definitions of functional urban areas (FUAs) are contiguous in this case, and subject to changes. We namely present the FUA definitions used for the 2006 and 2012 editions of the Urban Atlas land use dataset. The Lille area is multicentric, since 6 contiguous urban areas are identified by the Urban Atlas 2012 dataset. The Aix-Marseille area is duocentric, the 2 cities being merged in the 2012 version of the data.

In general, cities which are considered good examples of polycentrism in the literature fall into two groups.
First, very large cities, like Paris (Figure \ref{paris}) or London in Europe, with a clear main center, and where subcenters (smaller cities not far away) are also large, and appear more clearly than for smaller cities. In larger cities, one can argue that substructures are also larger, and appear more evidently, thus constituting secondary centers.
Second, very dense urban region, composed of cities of rather similar sizes and close to each other, which have begun to merge through urban sprawl. European examples include the Randstad in the Netherlands (Figure \ref{ranstad} -- see also \cite{broitman2020} -- which covers Amsterdam, Rotterdam, Utrecht and the Hague), the Rhine-Ruhr (Figure \ref{rhine-ruhr}, covering Essen, Dortmund, Wuppertal, Düsseldorf, Mönchengladbach, Köln and Bonn) and Rhine-Main regions (Frankfurt am Main, Darmstadt, Mainz, Wiesbaden) in Germany, the Lille or Marseille areas in France (Figures \ref{lille} and \ref{marseille}).

Examples of European polycentric urban regions mentioned in the literature (for instance \cite{hall2006}) also include Greater Dublin and Central Belgium (Brussels, Antwerp, Namur, Ghent, Bruges, Charleroi, Liège), including an area sometimes named the Flemish Diamond.
We also add in this group the Central England conurbation (Birmingham, Manchester, Liverpool, Leeds, Sheffield, Leicester, Stoke on Trent, Nottingham, Worcester, Coventry, Lincoln).

All these cities are considered candidates to polycentricity, and marked as such on the map of Figure \ref{europe}, where the overlap between different urban areas in polycentric regions is apparent.
Note that quite many of those polycentric areas are located in the ``Blue banana'' \cite{brunet1989} megalopolis, a highly urbanised area stretching from Liverpool to Milan.

We see from the maps that in Europe at least, polycentric urban areas are, 
as stated by \cite{kloosterman2001}, ``urban configurations where a number of historically distinct cities are within contemporary commuting distance''. This is especially clear on Figures \ref{lille},\ref{marseille} and \ref{paris}, where we can see that contemporary cities are extensions, through urban expansion, of cities already present in the 18th century.

\section{Models} \label{models}

After a first qualitative look at polycentric urban areas, we study them more quantitatively in this section. We first try to formalize more precisely the monocentric and polycentric city frameworks -- they remain rather theoretical and unrealistic. For instance, what would be a rigorously monocentric city? In theory, it should have a perfect circular symmetry (around the punctual city center) of the road network, of land uses, water bodies, population density, etc. This is obviously impossible. It would be quite interesting actually (but outside the scope of this chapter) to find which cities are closest to perfect monocentricity. One can however imagine a city which is very convincingly monocentric. Anyway, the monocentric city is rather a theoretical framework which does not easily translate to reality. So we will use it in a pragmatic manner and wonder if many or most cities can be reasonably considered monocentric.
Note that monocentric cities can be considered a special case of polycentric cities with only one center. The reverse does not work, obviously.

\begin{description}[Monocentric city]
\item[Monocentric city]{A city which clearly dominates its surroundings, without neighbours of comparable size. Secondary centers are small compared to the main center.}
\end{description}


Something quite similar can be said of polycentric urban areas (for a different reason). It is rather difficult to find an urban area which is convincingly polycentric. What would this be? A first answer would be simply an urban area which does not have a strictly circular organisation. It should have several centers, of similar or different sizes, which break the circular symmetry. In a strict sense however, this is true of all cities, as argued previously. In a strict sense, all cities are polycentric and no city is monocentric. This is quite close to the general opinion expressed in the scientific literature on the subject: cities are polycentric, not monocentric. However, distance to one main center remains the first spatial differentiation, in most urban areas, and radial profiles of different variables show surprising regularities \cite{lemoy2020,lemoy2021}, as developed in section \ref{radial}. Moreover, if all cities are polycentric, then the notion of polycentricity is not informative since it is not a discriminating criterion. Some cities must be more monocentric or polycentric than others. What would be a convincingly polycentric city then? We show some classical examples on the maps of Figures \ref{lille}, \ref{marseille}, \ref{ranstad}, \ref{rhine-ruhr} and \ref{paris}. But in those examples, polycentricity is rather found at the scale of a region comprising several individual cities, which might be rather monocentric on their own. These neighbouring cities have grown and expanded over long periods of time so as to overlap. The case where these centers have emerged through a rather recent history would be more convincingly polycentric, in the sense that it would not correspond to clearly separated neighbouring cities a few centuries ago which have expanded and recently collided. But we do not find examples of this type in our European dataset.

We find two different cases of polycentric areas though, represented in our examples: the case where a clear dominant center exists (Lille, Marseille, Paris on Figures \ref{lille}, \ref{marseille}, \ref{paris}), and the case where individual cities of rather similar sizes are involved (Ranstad or Rhine-Ruhr on Figures \ref{ranstad} and \ref{rhine-ruhr}).

\begin{description}[Polycentric urban region]
\item[Polycentric urban region]{A region where several different individual cities in proximity have started to merge through urban expansion and sprawl.}
\end{description}

Consequently, instead of finding very clearly monocentric or polycentric cities, we can define clear monocentric or polycentric perspectives on cities.

\begin{description}[Monocentric perspective]
\item[Monocentric perspective]{Studies cities in a monocentric (radial) manner, focusing on the center-periphery evolution of variables.}
\item[Polycentric perspective]{Studies cities in a polycentric manner, starting by identifying and counting centers and subcenters.}
\end{description}


We now turn to a quantitative analysis of empirical data regarding these polycentric cities, using first a radial or monocentric analysis, and then a short review of polycentric analyses.



\subsection{Radial analysis}
\label{radial}

\begin{figure}[!h]
\begin{center}
\begin{tabular}{cc}
\includegraphics[width=0.49 \textwidth]{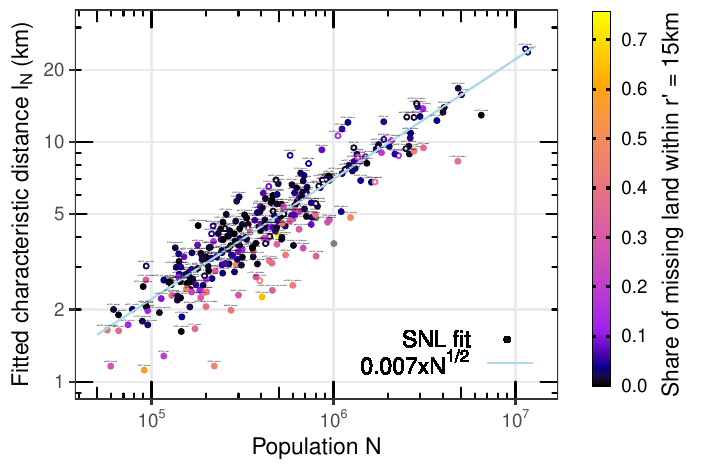} &
\includegraphics[width=0.49 \textwidth]{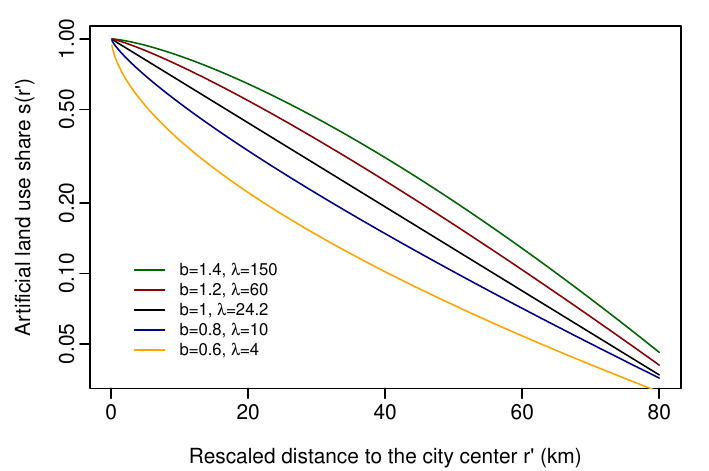}\\
\includegraphics[width=0.49 \textwidth]{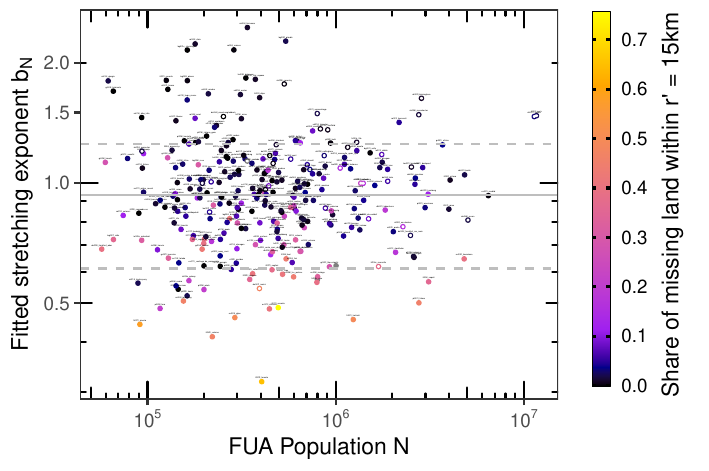} &
\includegraphics[width=0.49 \textwidth]{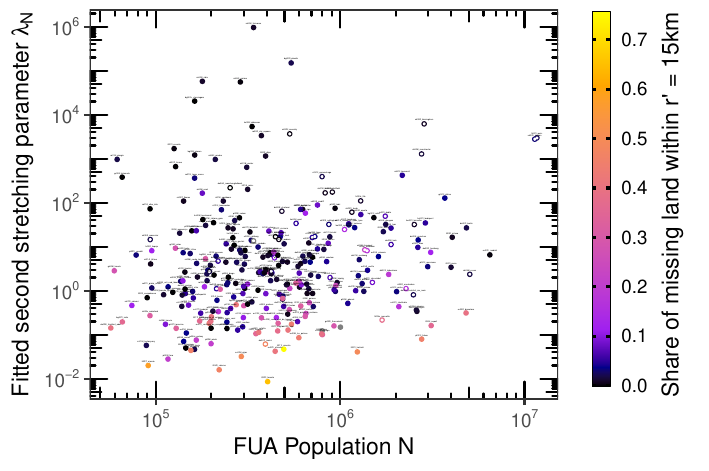}
\end{tabular}
\caption{Estimated parameters of the exponential (top left) and stretched exponential (bottom) fits of artificial land use profiles, displayed as functions of the total population $N$: characteristic distance $l_N$ (top left), stretching exponent $b_N$ (bottom left) and second stretching parameter $\lambda_N$ (bottom right). Cities which might correspond to polycentric urban regions are marked by a white dot. Top right panel: examples of stretched exponential profiles on a semi-logarithmic graph, with different values of the stretching parameters $b$ and $\lambda_N$.}
\label{fits}
\end{center}
\end{figure}

Many urban phenomena have been studied radially, including of course population density since \cite{clark1951}, land use (\cite{guerois2008},\cite{lemoy2020}), but also building height, air pollution, etc. In previous work, we focus mainly on a radial analysis of urban land use \cite{lemoy2021}, the variable which seems to present most regularity. We use the Urban Atlas 2012 dataset provided by the Copernicus Land Monitoring Service and study only the 300 urban areas which are available both for the 2006 and 2012 versions of the dataset. We show that the radial profile of land use in these urban areas evolves in a very consistent manner from the center (which presents mostly artificial land uses) to the periphery (with mostly natural land uses). More precisely, the share of artificial land $s(r)$ decreases exponentially with the distance to the center $r$. We use the (main or historical) city hall as the location of the city center, an area where land use is completely artificial, so that $s(0)=1$. Larger cities, in terms of their population $N$, have a larger extent in terms of artificial land use, which means that their radial profiles of artificial land use $s_N(r)$ decrease more slowly with distance $r$ to the center than in smaller cities. The radial profile has on average the mathematical form $s_N(r) = \exp(-r/l_N)$, where $l_N$ is the characteristic distance of this exponentially decreasing profile. We fit individual cities' radial profiles using this simple non-linear (SNL) fit (see Figure \ref{fits}) with a non-linear least squares method.

We also show that this characteristic distance $l_N$ scales like the square root of total city population $N$. On average, $l_N=l_1 \sqrt{N}$, with $l_1\simeq 7$m the characteristic distance of a (theoretical) city having one inhabitant \cite{lemoy2021}. This is illustrated on the top panel of Figure \ref{fits}, and corresponds to a homothetic scaling of radial land use profiles (in the two horizontal physical dimensions), where population $N$ is proportional to artificial surface $S_N \sim \pi l_N^2 \sim N $, since the characteristic radius $l_N$ is proportional to the square root of population, $l_N \sim \sqrt{N}$. This means that the artificial area per capita is constant and independent of the size of cities. 
This leads us also to defining a rescaled distance to the center $r'=rk_N$. The rescaling parameter $k_N=\sqrt{N_{\text{Paris}}/N}$ is needed to stretch the radial land use profile $s_N(r)$ of a city of total population $N$ to make it comparable to the profile of Paris, the largest urban area of the dataset with population $N_{\text{Paris}}$ (which we use as a reference, without loss of generality). For a city 4 times smaller in terms of population than Paris, like Milano, $k=2$. Note that this rescaled distance is used for the buffers of Figures \ref{lille}, \ref{marseille}, \ref{ranstad}, \ref{rhine-ruhr} and \ref{paris}, which consequently present comparable shares of urban land uses (see also \cite{tobler1969} for a quite similar approach).

When we study polycentric cities with this radial approach, we notice, on the top panel of Figure \ref{fits}, that the polycentric cities mentioned in section \ref{empirics} are not really outliers in this analysis. They follow rather faithfully the general tendency. This is surprising, since one could expect polycentric cities to stand out in a radial (monocentric) analysis. Indeed, polycentricity should break the circular symmetry of cities and yield unusual radial profiles.

To go further, we define an Urban Land Index ULI$=l_N^2/N$, which gives a measure of the total artificial area (in square meters) per inhabitant. This ULI is independant of city size $N$ and has an average of nearly 50 m$^2$. It is directly related to the residual of the model presented on the top panel of Figure \ref{fits}. Its value is actually slightly higher on average for our polycentric candidates cities than for all cities of the database. Its mean value is 49.5 m$^2$ (resp. 60.1 m$^2$) for all cities (resp. polycentric candidates), and 54.4 m$^2$ (resp. 62.5 m$^2$) when only ``continental'' cities are considered. Indeed, since coastal cities typically have low values of this Urban Land Index because of the area covered by water, we try to set aside the cities which have more than 20\% missing area (usually covered by water) within their first $r'=15$km from the center (that is, 15km at the scale of Paris).

We see with these values that our polycentric candidates actually stand out from the rest, but not to the point of being outliers, as can be seen on Figure \ref{fits}. They have higher values of the ULI on average, which is quite understandable since they are located in very urbanised areas, which provides them with more urban land, especially in their periphery, and a higher value of the characteristic distance $l_N$, compared to other cities of similar population.

\begin{figure}[!h]
\begin{center}
\includegraphics[width=\textwidth]{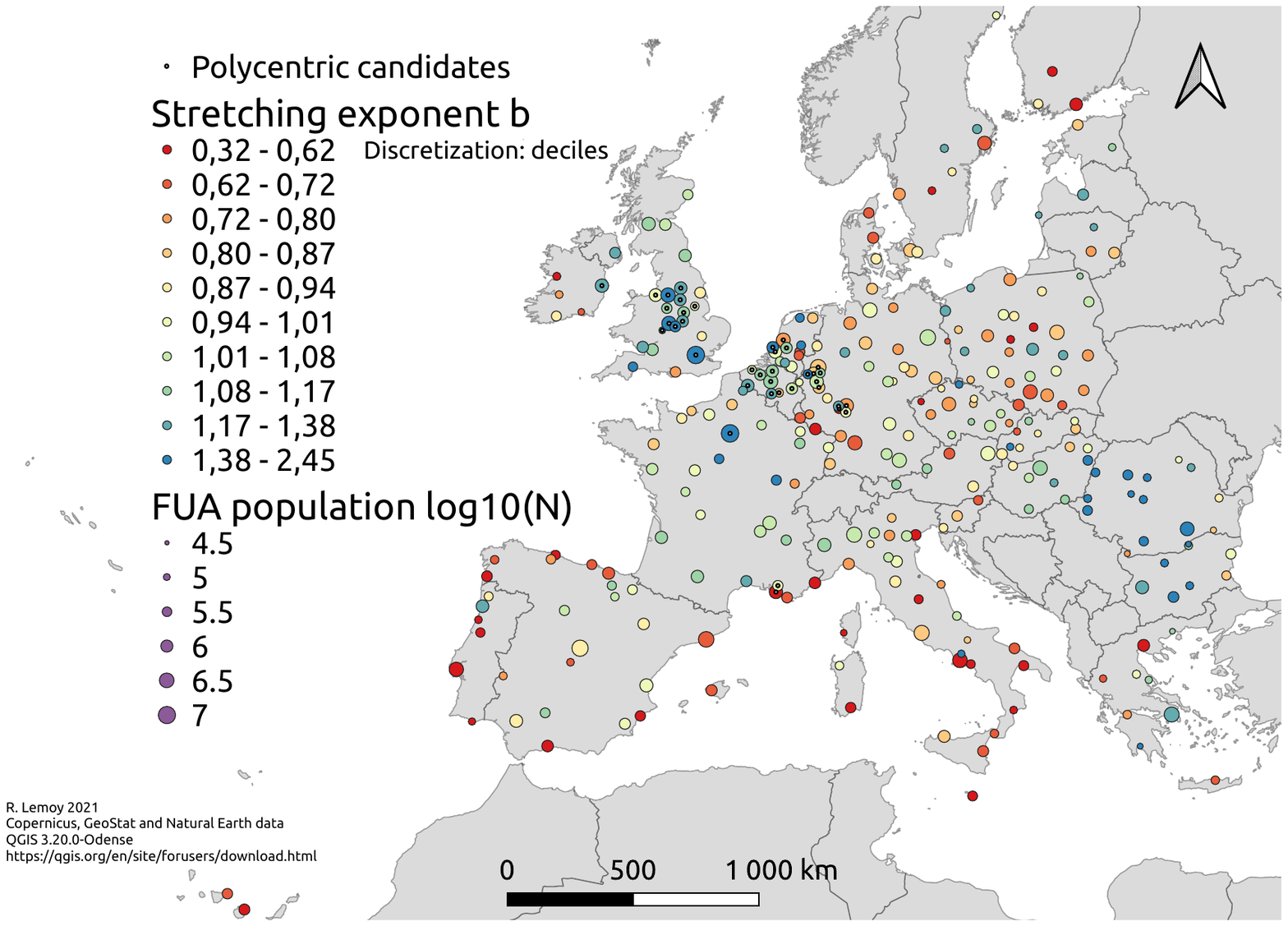}
\caption{Map of the stretching exponent $b$ of a stretched exponential fit of artificial land use radial profiles for large European urban areas in 2012, showing a rather clear national effect, low values for coastal cities (as expected) and rather large values for cities which are part of polycentric areas. The size of symbols indicates the decimal logarithm of FUA population $N$. The map is made using Copernicus, GeoStat and Natural Earth data with \href{https://qgis.org/en/site/forusers/download.html}{QGIS 3.20.0-Odense}.}
\label{europe}
\end{center}
\end{figure}

We also try to see how much urban areas deviate from the exponentially decreasing radial profile which we observe on average. To this end, we perform a stretched exponential fit of the form
\[
s_N(r) \sim \exp(-r^b/\lambda_N),
\]
where $b$ is the stretching exponent and $\lambda_N$ the second fitting parameter. Compared to an exponential function where $b=1$, values higher than one, $b>1$, indicate a more concave profile, while values smaller than one, $b<1$, correspond to a more convex profile. $\lambda_N$ is a kind of characteristic distance, but which actually only has the dimension of a distance for $b=1$. The results are shown in Figure \ref{fits}, and are also mapped on Figure \ref{europe}. We observe that the stretching exponent is nicely distributed around $b=1$, corresponding to an exponential profile, in a lognormal manner (see \cite{lemoy2021}). Polycentric candidates have clearly higher values of $b$ on average, which means that their radial profile is more convex than the average. This goes with higher values of the second fitting parameter $\lambda_N$ also. 

This can be understood with the top right panel of Figure \ref{fits}, which displays the shape of stretched exponential radial profiles for different values of $b$ and $\lambda_N$, including the average exponential profile ($b=1$), which appears as a straight line on this semi-logarithmic graph. Indeed, polycentric cities have a higher value of the stretching exponent $b>1$, which indicates that the share of artificial land is higher in areas close to the city center, compared to cities with no close neighbours. A higher value of the second fitting parameter $\lambda_N$ is also needed to compensate the higher value of $b$, as shown by the top right panel of Figure  \ref{fits}. On the contrary, we can note that coastal cities have low values of $b<1$, and accordingly low values of $\lambda_N$. This can be linked to the fact that part of the land is unavailable for artificialization, because it is covered by water. This yields a smaller share of artificial land in central areas of coastal cities, contrary to cities which are part of polycentric areas. We also note that cities in Romania, Bulgaria and the United Kingdom have especially high values of the stretching exponent $b$.

\subsection{Intersection of two neighboring cities}

What can we expect in theory to appear at the intersection of two neighbouring cities, when studied with a radial analysis? We note in this subsection that at least three different kinds of phenomena could be expected. These phenomena depend on the studied variables, and we propose some examples of variables which could display them. Such situations could be studied in more detail in future work using pertinent empirical data, in order to gain some insights on the inner workings of polycentric regions.

The first example is that of housing price. In an area between two neighbouring cities, one can expect that the housing price in a given location will be close to the maximum of the price one could expect from the proximity to either of the two considered cities \cite{laziou2023}, because such an area provides at least the urban opportunities offered by the city with more influence (in terms of size and distance). The same can be expected of urban land uses, and maybe also of population density. Let us note also that \cite{jiao2015} performs radial and polycentric analyses of urban land use based on this hypothesis.

One can consider rather the transport cost (in terms of time or distance) to the nearest city center, where job opportunities are concentrated. This variable is of crucial importance in urban economic theory, and studied empirically by \cite{mennicken2022}. In this case, in the area where two close cities overlap, the transport cost (distance, time, monetary cost) to the nearest center is the minimum of the costs of trips to either of the two neighbouring city centers.

A third example is given by air pollution, for instance the concentration of nitrogen dioxide NO$_2$, as measured by satellite as an average over a tropospheric column \cite{schindler2016}. This air pollution is mainly caused by traffic, through combustion of fuel. As such, it is concentrated in urban areas, but it disperses over regional scales. In this case, in the considered overlapping area of two close cities, one can expect the air pollution to be governed by the sum of the concentrations generated by both cities. A summary is given in Table \ref{join}.

\begin{table}[!h]
\begin{center}
\begin{tabular}{p{0.39\textwidth}|p{0.12\textwidth}|p{0.39\textwidth}}
Variable & Intersection & Justification\\
\hline
Housing price, urban land uses, population density & Maximum & Non-additive phenomenon\\
Transport cost to the center & Minimum & Adding centers adds opportunities and increases accessibility\\
Air pollution & Sum & Additive phenomenon
\end{tabular}
\label{join}
\end{center}
\caption{Interactions at the intersection between two neighbouring urban areas.}
\end{table}

\subsection{Polycentric analysis}

We focus here on a radial analysis of polycentric cities, since we use this approach repeatedly in previous research. 
Indeed, a monocentric city is quite easy to define. The location of the city center is not a very hard problem. For instance, the location of the main or historical city hall can be chosen \cite{wilson2012, walker2016,lemoy2020}. The definition of a polycentric city is much more complex. Once polycentricity is assumed, one needs to define centers and subcenters in empirical settings. This problem is already hard enough to remain quite open after a few decades of research on polycentric cities.
There are many competing definitions of centers and subcenters, and methods to locate them.

Thus, a body of scientific literature was built on the analysis of polycentric cities directly. This 	multidisciplinary literature focuses in particular on the identification and count of centers and subcenters through different techniques. In this strand of literature, radial analysis is usually not considered. The polycentric structure of cities is a preliminary assumption. We review several works which are presumably representative of the different approaches developed in different disciplines.
Studies start in economics, where centers are associated to employment centers, following the standard urban economic model. \cite{mcdonald1987} defines centers using cutoffs on employment density and employment population ratio. This approach with cutoffs is quite wide-spread and used by subsequent literature. For instance, 
\cite{giuliano1991} define centers as contiguous ``zones'' (transportation analysis zones) of sufficiently high employment density and total employment. A drawback of using cutoffs is the introductions of rather arbitrary parameter values, and a part of the literature, in different disciplines at the crossroads of an emerging science of cities, strives to develop non-parametric methods.
In an urban area having one main center (CBD), \cite{mcmillen2001}, and other works, identify subcenters as significant residuals of the employment density radial profile. This profile has an exponential form, and they look for residuals of a locally/geographically weighted regression of logarithmic employment density (other works use different regression methods).
The theoretical model of reference in economics in this case is given by \cite{fujitaogawa}, who propose a rather complex extension of the standard urban economic (AMM) model, which predicts, regarding polycentricity, that the number of subcenters grows with population size -- a quite understandable stylized fact, which is verified empirically.
In geography, \cite{riguelle2007} study local autocorrelation of logarithmic employment density in Belgium and find that polycentrism is weak. \cite{arribas2014} do a related work on US cities at different times, without logarithm.
Note that this research in economics, geography and regional science is ongoing, proposing new measures of polycentricity.

More recently, a literature emerged in complex systems with methods sometimes inspired by statistical physics, using new datasets generated by digital footprint.
\cite{roth2011} identify centers in London based on real-time public transport data.
\cite{louail2014} find hotspots of human activities using mobile phone data in Spain. Their nonparametric method yields a number of subcenters which scales sublinearly with city size, in accordance with a related theoretical model \cite{louf2013}.
This rich literature is interesting in order to delineate centers and subcenters, but still leaves room for more empirical work on the question of monocentric versus polycentric cities. Linking these (empirical) polycentric approaches with radial ones in order to obtain new insights on cities and systems of cities would be an interesting research perspective, which is rather unexplored to our knowledge.


\section{Criticism}

Despite discussions in the scientific literature \cite{bailey2001, meijers2008}, polycentricity has become a normative concept in urban planning, in particular in Europe \cite{esdp1999,espon2005}, but also on other continents. Indeed, it is supposed to be associated with increased economic outputs related to agglomeration economies (see \textcolor{black}{Chapter 12. Introduction to agglomeration economics}). This is also linked with the positive literature about polycentrism from the 1970's - 80's \cite{schneider1981,gordon1985}, where this phenomenon is seen as an efficient way to develop public transportation, with a system of hubs and spokes associated with centers and subcenters. Polycentricity is also seen as a desirable, decentralized model of governance.

However, if in Europe polycentric regions are mostly cities sprawling and merging, as we saw from this study, this policy goal resembles rather a desire to steer the larger, more fundamental and unavoidable process of urban expansion and sprawl. Then, this issue of urban sprawl should probably rather be tackled directly. It is associated with negative outcomes such as losses of arable land, of biodiversity, air pollution, climate change linked to emissions of greenhouse gases, economic vulnerability of periurban households which are dependant on cars and oil-based fuels. Note that this is quite unsettling since polycentricity itself, which we saw here as directly related to urban sprawl, is seen as a positive policy goal. Urban sprawl is usually explained, in particular in the economic literature, by the decreasing price of transport, in particular commuting to work (which is concentrated around city centers).

With this, the question of polycentricity comes back to the much debated question of transport costs. If one considers benefits over the long term (a few decades or centuries), transport and commuting costs are probably too small nowadays, despite recent increases in fuel prices. For instance, it is probably not worth it right now to use a car instead of a bicycle for short urban trips (of less than 3km for instance), when one takes into account the fact that people in a few generations will have a much more difficult access to those precious fossil fuels (like oil) on which our current economic activity is grounded -- in addition to living in a world with a quite different climate. We refer the reader to the discussion in \textcolor{black}{Chapter 3 Mobility}.

We should also mention that the word ``polycentric'' is used in different contexts related to urban planning, and especially in the context of governance, where it is seen as a positive goal (in line with decentralisation). For instance, a polycentric governance of the European Union (EU) -- at this scale, the Blue Banana region mentioned above is the center \cite{faludi2005}. This brings us to a discussion on spatial scales, since polycentricity is mentioned at the scale of the EU (continental scale), at the scale of a conurbation like the Rhine-Ruhr region (see Figure \ref{rhine-ruhr}), with Essen as a possible center (these are both inter-urban scales, the scales of systems of cities, \textcolor{black}{see Chapter 10}), and also at the intra-urban scale (the urban areas of Paris or London for instance). 

Are we then speaking of a polycentric city or of a system of cities in proximity? It is not easy to set the limit between both cases, and the urban planning literature is not always clear in this respect. Many studies are carried out at the inter-urban scale, but we have considered polycentricity here mainly at the intra-urban scale. This constitutes the major research issue in the domain, which is simply the robust definition of the system to be studied.

\section{Open issues}

Cities which are usually described as polycentric, and which we rather describe here as conurbations, are a real scientific challenge mainly when we want to isolate them, that is, in terms of their definition or delineation. Is there a main, central city to whose urban area the whole region could be considered to belong? If we are rather in a case where several urban areas are mixed, which city does this specific neighbourhood belong to? 

This question of city (or rather urban area) definition is a major research question and open problem in urban science, and polycentric urban regions are a central challenge in this respect. Municipalities (administrative definitions) are usually not appropriate for scientific studies since they are too arbitrary (really small and numerous in France for instance, larger in Germany). Functional Urban Areas (FUAs) are usually considered a good definition, but many other methods exist, such as morphological definitions (see \textcolor{black}{Chapter 1 and Chapter 8}).

\begin{figure}[!h]
\begin{center}
\includegraphics[width=\textwidth]{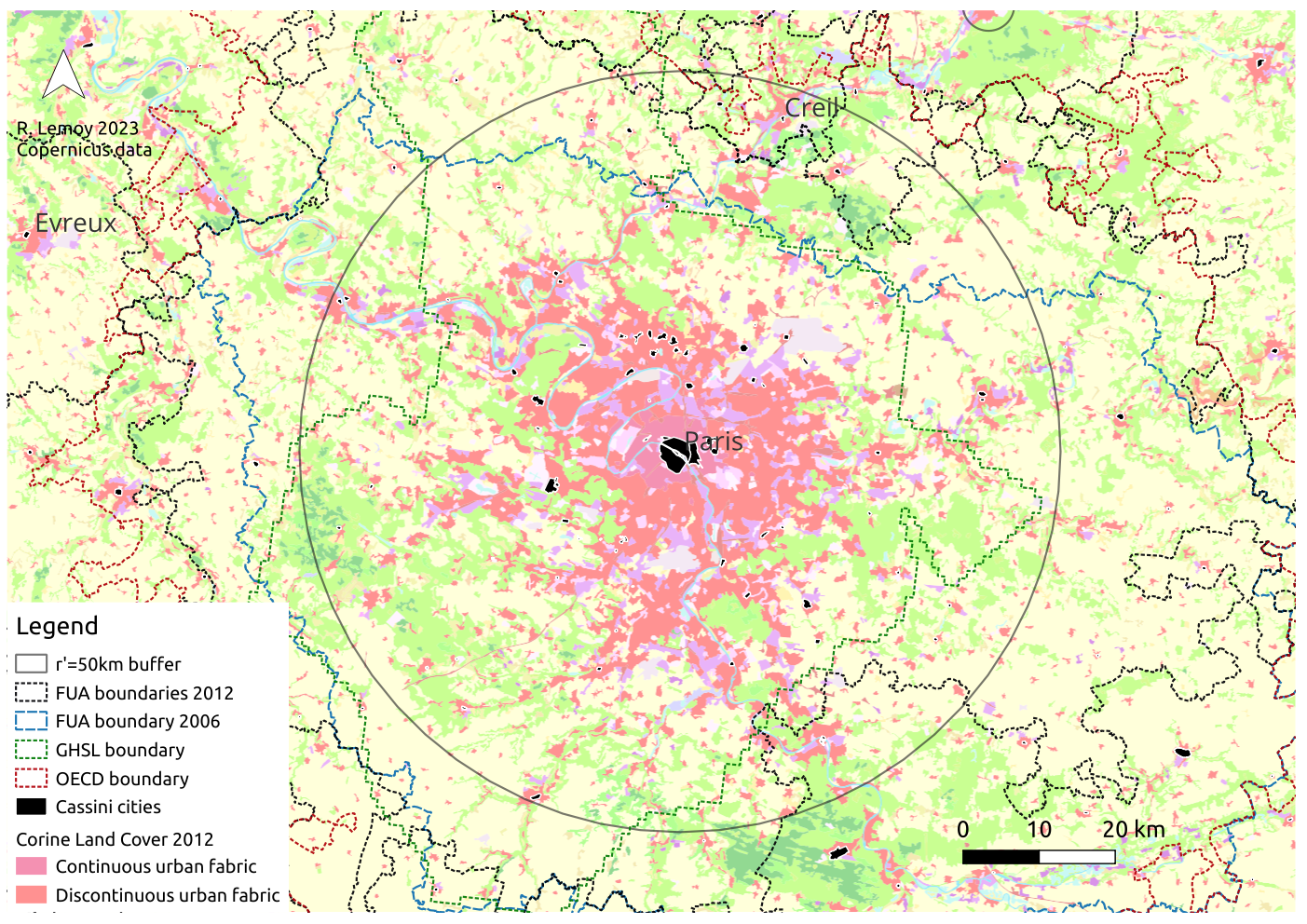}
\caption{Land use map of the Paris FUA (France). The standard Corine Land Cover color scheme is used. The buffer represents a radius $r'=50$km around the center. FUA boundaries are presented: Urban Atlas in its 2006 (blue) and 2012 (black) versions, GHSL (green) and OECD (red). Data from the Cassini map is overlaid, showing cities' extent in the 18th century.}
\label{paris}
\end{center}
\end{figure}

This issue is illustrated by Figure \ref{paris}, which presents the results of the definition of the Functional Urban Area (FUA) of Paris using different methods and data. The aim is the same in each case. The FUA is defined based on home-work commuting trips, which indicates the functional extent, in terms of employment area, of cities. However, the results differ significantly, especially in the peripheral parts of the urban area. We can also refer to Figure \ref{marseille}, where two distinct FUAs (Marseille and Aix-en-Provence) in the 2006 edition of the Urban Atlas dataset have been considered as just one in its 2012 edition. Or similarly, different extents of urban areas with the same datasets can be seen on Figure \ref{lille}.

This issue has of course many implications, as it influences most studies of cities, for instance the distribution of city sizes (Zipf law for cities, \textcolor{black}{see Chapter 9}). Should the secondary cities (centers) around a large city be counted as individual cities, or just be integrated in the large city? With the first option, the city size distribution has a shorter tail and more cities are counted in the bulk of small urban areas. In the second option, the tail tends to be longer with large cities becoming even larger. This has a direct influence on the exponent governing the power-law tail distribution.

In general, the question of the city definition is probably not enough considered in particular in 0-dimensional analyses of cities (see \textcolor{black}{ Part III of the book}), where definitions of urban areas from statistical institutes are usually taken as inputs and not really questioned -- we should still mention notable exceptions like \cite{cottineau2016}. For instance, all values of (urban) scaling exponents found in the literature (\textcolor{black}{see Chapter 11}) are dependant on this issue of polycentricity and city definition. Studies of cities using percolation techniques (\textcolor{black}{see Chapter 16}) also face the problem of polycentric urban regions, which blur the boundaries between individual cities.

We also mention that we mostly considered urban land use as the variable of interest in this chapter (as illustrated by different maps and analyses), but all urban variables are concerned by this issue of polycentricity. And land use happens to be the variable which has the highest spatial scaling exponent (0.5 or 1/2) among the variables studied with radial and scaling analysis so far \cite{lemoy2020}. Population density has a smaller exponent, around 1/3, and housing price an even smaller one. As a consequence, the characteristic spatial scales of housing price and population density variations are notably larger than those of land use \cite{laziou2023}, and polycentricity is an even more serious issue when studying these variables. The influence of a small or medium-size city on housing price and population density is clearly further-reaching than on land use. Land use could actually be the best variable to delineate individual cities, which is rather well illustrated by the different maps presented in this chapter, where individual cities can be quite well distinguished even in clearly polycentric areas.

\section{Summary}

In this chapter, we explore monocentric and polycentric cities in Europe mainly using radial (center-periphery) analysis and spatial scaling laws.
We observe that polycentric regions are the result of the urban expansion and sprawl of cities which have been neighbours for centuries and start to merge recently. They can have one clear main center (London, Paris) or a less hierarchical distribution of centers (Ranstad, Rhine-Ruhr).

Their radial profiles of urban land use are not really exceptional compared to average cities, although these polycentric cities are on average more urbanized compared to cities of the same size. They also have a more concave radial profile of artificial land use, indicating that more land is urbanised, in the central areas of the city, compared to cities of the same size. This can be opposed to coastal cities, which tend to have more convex profiles because of the areas covered by water.

One key issue which scientists are facing concerning polycentric urban areas is the still open problem of city definition. It is indeed quite difficult to define an urban area in general (where should one set the limit? which criteria should be used?), and more particularly, to separate the individual cities which constitute a polycentric region.

This problem is currently gaining importance as more cities are expanding, sprawling and merging worldwide (\textcolor{black}{see Chapter 17}), constituting more polycentric urban regions. One can wonder how this will go on in the mid or long term. Is the future of urban areas made of ever more urban expansion, which would make this issue of polycentricity even more prominent? Or will cities face a contraction due to new factors (like rising prices of energy, non-renewable resources and transport)?

To conclude, all cities can be considered polycentric, but the radial or monocentric viewpoint is still quite interesting as a simple robust intra-urban model. If only one variable should be kept for a rough or parsimonious intra-urban analysis, very clearly distance to the center is the variable to keep. 
This discussion amounts largely to a question of viewpoint. When you look at urban areas from far away, you usually identify one main center. But if you look more closely, you see a polycentric structure.
The monocentric and polycentric formalisms of cities have been studied for decades already, and they are probably going to coexist for some time more.


%

\begin{acknowledgement}
The author acknowledges insightful discussions with Geoffrey Caruso and Gaëtan Laziou, the help of Walid Rabehi and Julien Perret for historical data, and helpful comments from Diego Rybski.
\end{acknowledgement}

\ethics{Competing Interests}{
This study was funded by ANR throught the GreenLand project and Normandy region through RIN project SUCHIES.}

\bibliography{biblio}
\bibliographystyle{unsrt}

\end{document}